\documentstyle[prl,twocolumn,aps,epsf]{revtex}
\begin{document}
\twocolumn[\hsize\textwidth\columnwidth\hsize\csname@twocolumnfalse%
\endcsname
\title{Partially filled stripes in the two dimensional Hubbard model:
statics and dynamics} 
\author{E. Louis$^{\ddag}$,
F. Guinea$^{\dag}$, 
M. P. L\'opez-Sancho$^{\dag}$ and J. A. Verg\'es$^{\dag}$}
\address{
$^{\ddag}$ Departamento de F{\'\i}sica Aplicada
and Unidad Asociada of the Consejo Superior de Investigaciones
Cient{\'\i}ficas (CSIC),
Universidad de Alicante, Apartado 99, E-03080 Alicante, Spain. \\
$^{\dag}${I}nstituto de Ciencia de Materiales de Madrid,
CSIC, Cantoblanco, E-28049 Madrid, Spain.}
\date{\today}
\maketitle
\begin{abstract}
The internal structure of stripes in the two dimensional
Hubbard model is studied by going beyond the Hartree-Fock
approximation. 
Partially filled stripes, consistent with experimental
observations, are stabilized by quantum fluctuations,
included through the Configuration Interaction method.
Hopping of short regions of the stripes in the
transverse direction is comparable to the bare
hopping element. The integrated value of $n_{\bf \vec{k}}$
compares well with experimental results.
\end{abstract}
]
\narrowtext
By now it is well established that charged stripes are 
formed in a significant doping range of 
cuprate oxides\cite{Cetal91,MAM92,Tetal95}.
The existence of these stripes was predicted, on the basis 
of mean field calculations, in advance of its 
observation, both in the one \cite{ZG89,PR89,Sc90,VL91} and the three
bands Hubbard model \cite{VG92}.
This is one of the scarce theoretical results in the
field of high-T$_{\rm c}$ superconductivity which was
confirmed after its prediction. Interest in these calculations
decreased, as it was generally understood that the Hartree-Fock
approximation was unable to obtain the partially filled
stripes observed experimentally. Unrelated calculations
found stripes with different fillings in the t-J model\cite{WS98},
and, using different techniques, in the Hubbard model\cite{Fetal99}, although
other numerical calculations show conflicting results\cite{HM99,CS99}.
Alternatively, it has been argued that stripes in doped
Mott antiferromagnets arise from  a tendency towards phase
separation, frustrated by electrostatic interactions\cite{EK96,Cetal98}.
Stripes in the Hubbard model have also been analyzed
within slave-boson techniques, which use the Hartree-Fock
solutions as input\cite{SSH98}.

In the present work we show that the mean field calculations, initially 
used to demonstrate
the existence of stripes, can be systematically improved in order to 
study the partially filled stripes observed experimentally.
In addition, they provide significant insight into the internal
structure of the stripes and their fluctuations in the transverse
direction. The Hartree-Fock method can be considered a quasiclassical
approximation to the spin and charge degrees of freedom.
Their low amplitude quantum fluctuations can be incorporated by using
the Random Phase Approximation\cite{GLV92}. In addition, one 
needs to consider quantum tunneling processes between 
degenerate, or nearly degenerate, Hartree-Fock solutions, 
when there are many. This is achieved with the
Configuration Interaction method (CI), widely used in 
quantum chemistry\cite{LC93,LG98}. The combination of the Hartree-Fock
and CI methods gives reasonable results even when applied to
one dimensional systems\cite{BJ00}. The CI method
restores the symmetries broken by the Hartree-Fock approximation,
and provides information on the quantum dynamics of the
static solutions obtained in mean field, which can
be broadly classified into spin polarons or stripes.

Previous mean field studies have focused on filled stripes\cite{VL91},
which tend to be the solutions with the lowest energy per hole
in this approximation, especially for the
values of $U/t \sim 4$ used in the initial studies\cite{ZG89,PR89,Sc90,VL91}.
There are, however, self consistent
solutions which describe partially filled stripes. These solutions
have energies per hole comparable to those of the filled stripes
for $U/t \sim 8 - 20$. 

\begin{figure}
\begin{center}
\mbox{\epsfxsize 5cm \epsfbox{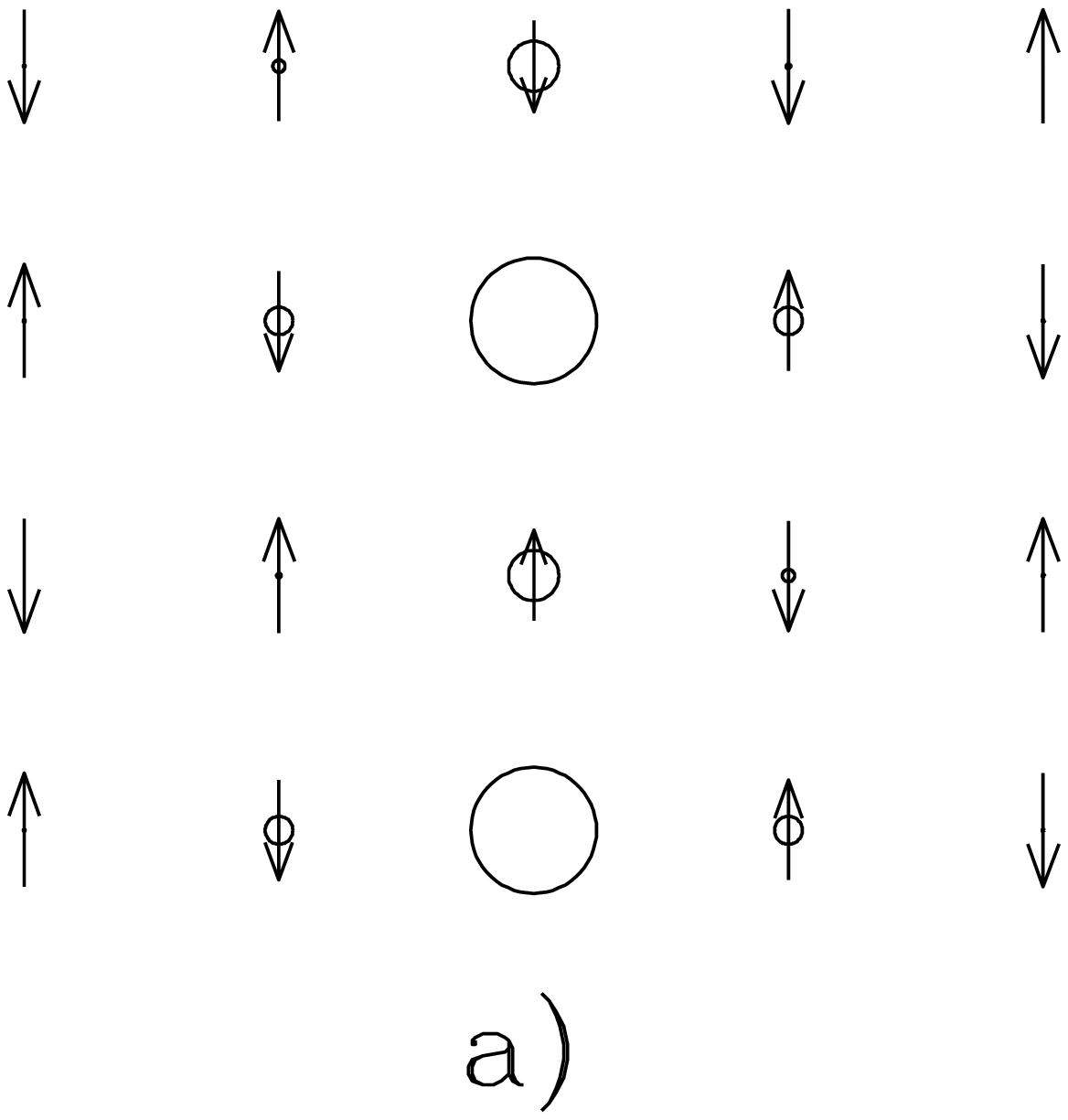}}
\mbox{\epsfxsize 5cm \epsfbox{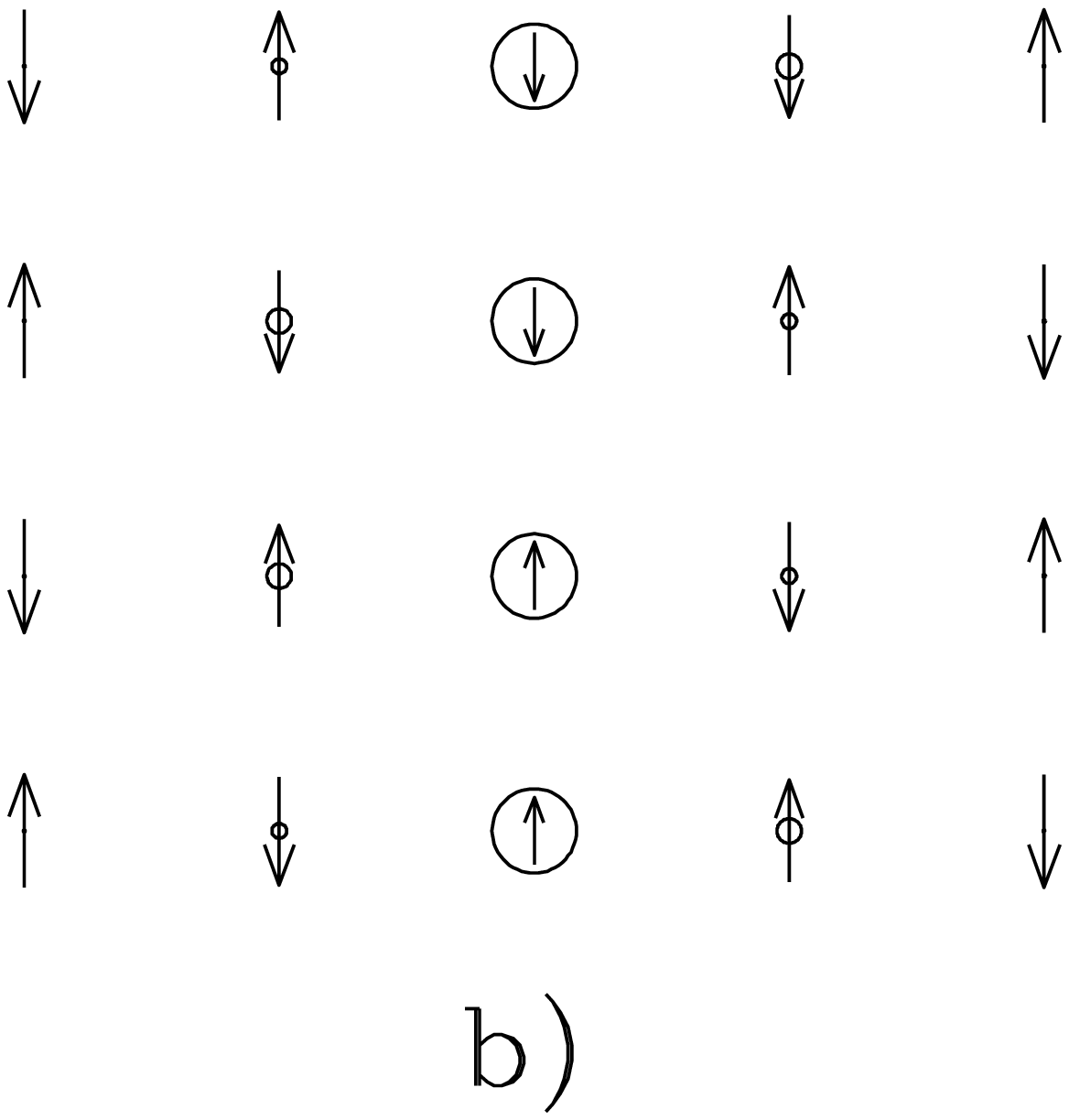}}
\mbox{\epsfxsize 5cm \epsfbox{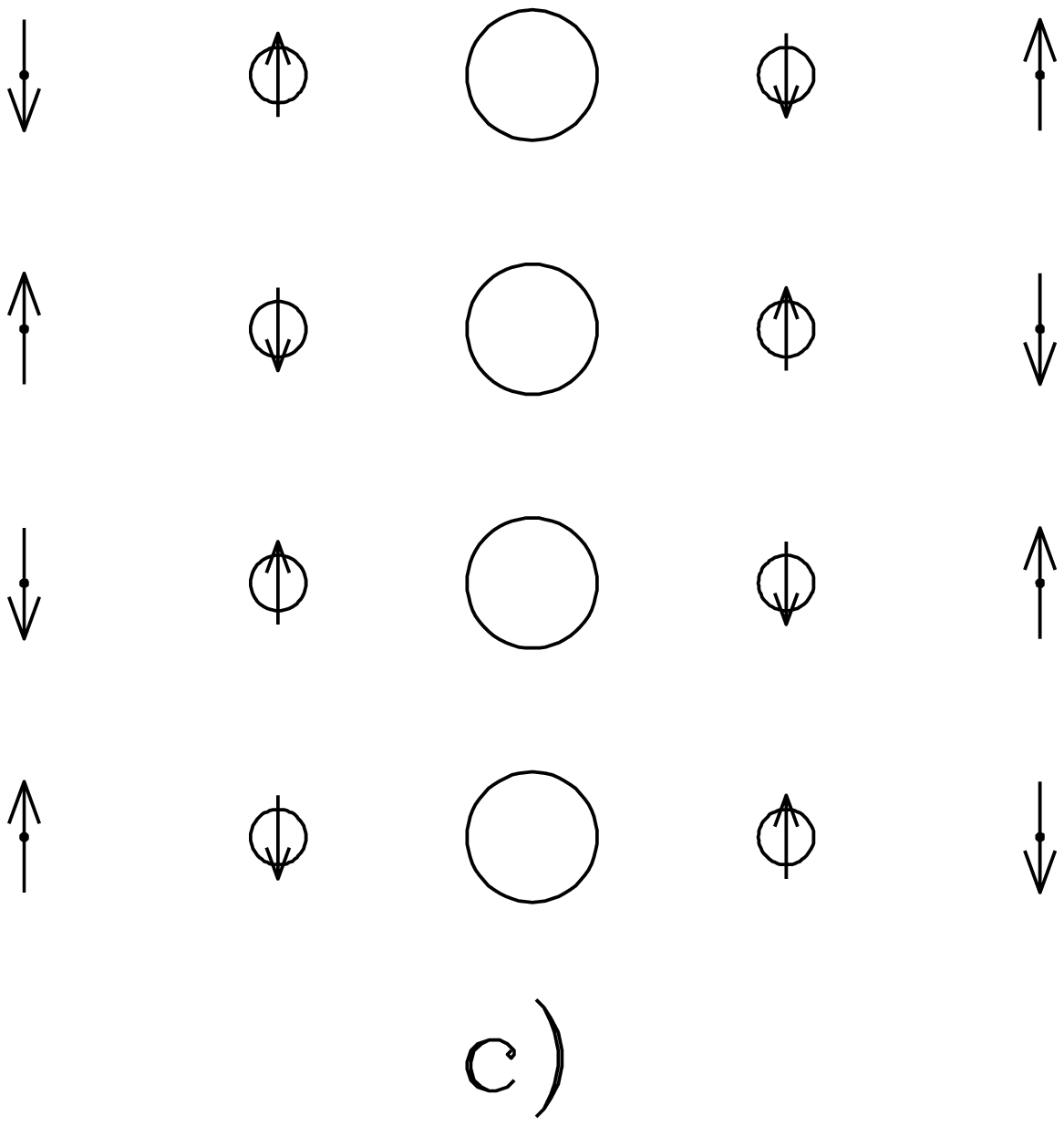}}
\end{center}
\caption{Spin and charge textures for the three stripe solutions discussed
in the text. Solutions a) and b) correspond to hal filled stripes,
while solution c) is a filled stripe.}
\label{fig:textures}
\end{figure}

\begin{figure}
\begin{center}
\mbox{\epsfxsize 8cm \epsfbox{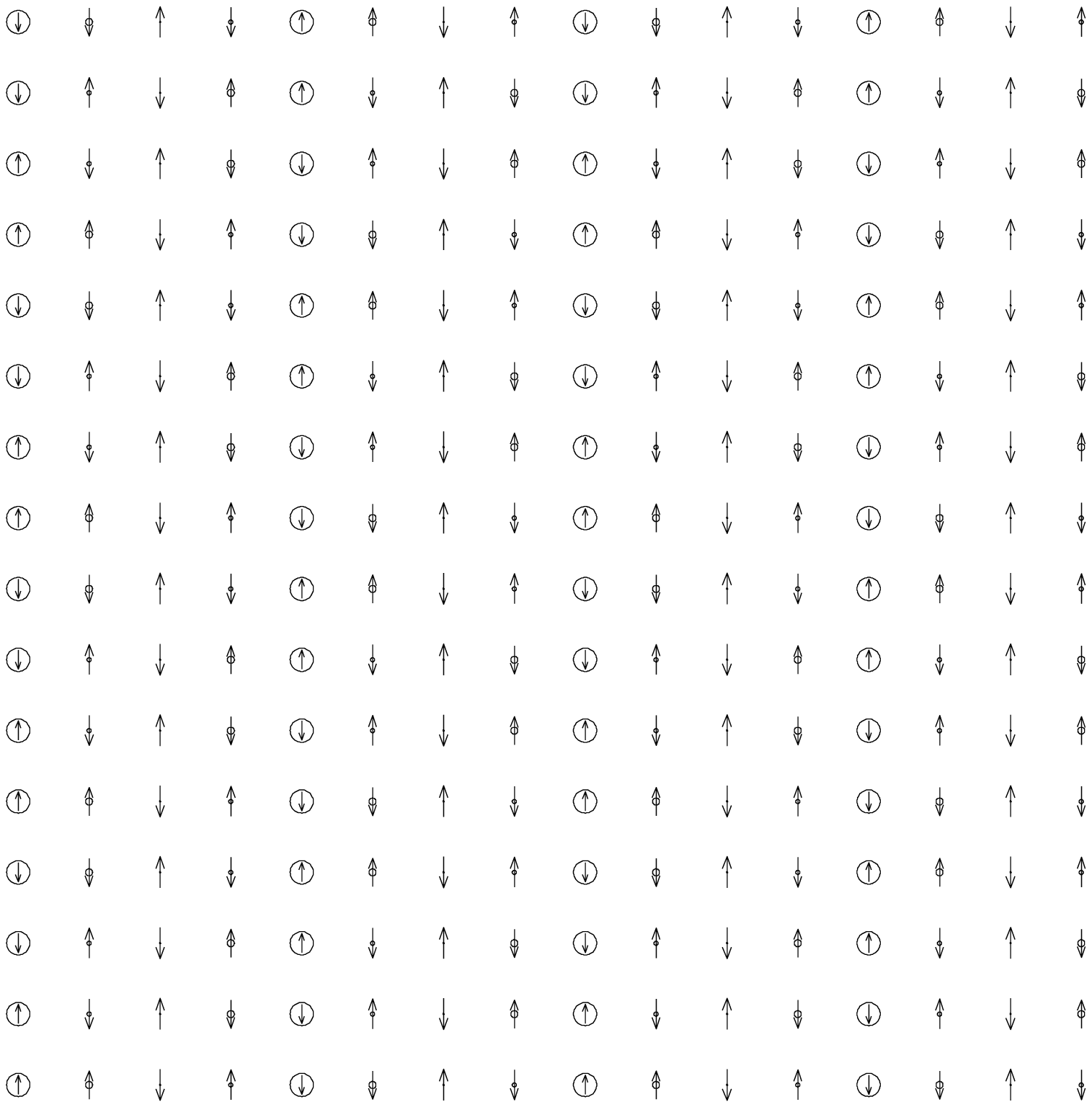}}
\end{center}
\caption{Spin and charge textures for a $16 \times 16$ cluster with 32
holes ($x$=1/8, where $x$ is the number of
holes per site ) and $U/t = 8$.}
\label{fig:cluster16}
\end{figure}

Typical stripe solutions obtained in this work within the mean field 
approximation and  periodic boundary conditions, are shown in
Fig.[\ref{fig:textures}], where, for simplicity, we show results
obtained in a $5 \times 4$ cluster, and $U = 8$ (hereafter we shall use 
$t$ as the unit of energy). The filled stripe, like the ones
considered in previous mean field calculations, is shown in
Fig.[\ref{fig:textures}] c), while Fig.[\ref{fig:textures}] a) and b)
correspond to half filled stripes, not reported earlier. These textures
correspond to self consistent Hartree Fock solutions, and can be found
in clusters of any size. A generalization of the solution shown in
Fig.[\ref{fig:textures}] b) to a $16 \times 16$ cluster with 1/8
holes per site (32 holes) is shown in Fig.[\ref{fig:cluster16}]. 
We have also calculated the energy required to create a hole in these 
configurations. In order to subtract the energy of the antiferromagnetic 
background we have averaged the energy per site obtained on 
$4 \times 4$, $4 \times 6$ and $6 \times 6$ clusters, the result is  
0.463 per site. The so obtained energies per hole are reported in Table I.  
It is interesting to note that the standard solution for the filled stripe 
shown in Fig.[\ref{fig:textures}] c) is the best UHF solution only for 
sufficiently small $U$,
while at intermediate $U$ (including the case of $U=8$ discussed here), 
there are better solutions which are inhomogeneous along the
stripe direction, with $\epsilon_{HF} = -1.081 t$, slightly smaller
than the energies of the half-filled stripes.
We also show in Table I  the energy gain due
to hybridization with displaced solutions of the same type
($\epsilon^0_{CI}$).

\vfil\eject
\twocolumn[\hsize\textwidth\columnwidth\hsize\csname@twocolumnfalse%
\endcsname
]
\narrowtext
\begin{tabular}{|l|l|l|}
\hline
Half filled stripe
&Half filled stripe
&Filled stripe \\
Fig.[1] a)
&Fig.[1] b)
&Fig.[1] c) \\
\hline
$\epsilon_{HF} = -0.959 t$
&$\epsilon_{HF} = -0.991 t$
&$\epsilon_{HF} = -0.958 t$ \\ \hline
\multicolumn{2}{|c|}{$\epsilon^0_{CI} = -1.47 t$}
&\multicolumn{1}{l|}{$\epsilon^0_{CI}= -1.12 t$} \\ 
\multicolumn{2}{|c|}{$\epsilon_{CI} = -1.57 t$}
&\multicolumn{1}{l|}{$\epsilon_{CI}= -1.36 t$} \\ \hline
\end{tabular}
Table I. Energies, per hole, of the stripe solutions shown
in Fig.[1] \\ \\
The best energies per hole
that we have obtained, after combining, within the CI scheme,
the previous configurations and others not shown in
Fig.[\ref{fig:textures}] are $\epsilon_{CI} = -1.573t$ (half
filled stripes) and $\epsilon_{CI} = -1.364 t$ (filled stripes).

The results in Table I  have been obtained for a single stripe.
They have uncertainities
due to the cluster size dependence and the substraction of the
antiferromagnetic background. In general, the energy differences are small,
of a few hundredths of $t$.
The stabilization of half filled stripes seems to be
a general result, at least for values of $U/t$ in the range 6-20.
The value $U/t = 8$ corresponds to a t-J model with
$J/t = 1/2$. The energy per hole that we obtain compares
well with calculations for this model\cite{WS00}. For $U/t = 12$
($J/t = 1/3$), we obtain $\epsilon_h \approx -1.71 \pm 0.05 t$.
This value is
also in reasonable agreement with other calculations\cite{WS98b}.

We have also calculated the hole--hole 
correlation, $C_{i,j}=\langle ( 1 - n_i ) ( 1 - n_j ) \rangle$ 
for the half-filled stripe in a 
$5 \times 4$ cluster. The optimal CI wavefunction, 
discussed above, was used in the calculation. $C_{i,j}$
shows a maximum (0.021) when the holes are separated by a
vector  ${\bf r}_{ij}={\bf r}_i-{\bf r}_j=(0,2)$, that is when they are
two lattice constants apart along the stripe direction.
This feature of the hole-hole correlation differentiates stripes from
other two hole configurations investigated within the $t-J$ \cite{RD97} or the
Hubbard models \cite{FO90,LG98}. We note, however, that $C_{i,j}$ is also rather
large (0.011) for the hole-hole separation at which the results 
reported in those studies show the maximum hole-hole correlation, namely,
${\bf r}_{ij}=(1,1)$.

Our method is well suited to analyze quantum fluctuations of stripes in the
transverse direction, as we can calculate the off diagonal terms of the
full hamiltonian between stripe solutions which are displaced by an arbitrary
lattice constant. We find that the matrix element which describes the 
shifting of a stripe segment of length $l$ 
by one lattice unit decays as $t_s e^{-l/l_0}$, where
$t_s \sim t$ and $l_0 \sim 2$.

Note that the two main configurations involved,
Fig.[\ref{fig:textures}] a) and b), can be viewed as a superposition of 
site centered and bond centered one dimensional domain walls\cite{BJ00}.
These configurations lead to a very accurate description of the
dynamics of a hole in the one dimensional Hubbard model\cite{BJ00}.
The fact that the 
maximum hole-hole correlation
is found along the stripe (see discussion above)
implies an additional similarity
between the stripe solution and one dimensional Luttinger liquids.
Thus, the stripes can be viewed as a one dimensional system 
at quarter filling. 
Because of the commensuration with the lattice,
we find a gap in the charge spectrum. We estimate the value of this
gap, $\Delta_{ch}$, beyond mean field theory,
by calculating $\Delta_{ch} = E_{n_h + 1}
+ E_{n_h - 1} - 2 E_{n_h}$, around the optimal filling.
We obtain, for $U/t = 8$, $\Delta_{ch} = 0.7 \pm 0.1 t$, which is lower
than the Hartree-Fock gap. For $U/t = 12$,
we find $\Delta_{ch} = 0.5 \pm 0.1 t$,
which compares well to the gap obtained using dynamical mean field
techniques\cite{Fetal99}. 

\begin{figure}
\begin{center}
\mbox{\epsfxsize 10cm \epsfbox{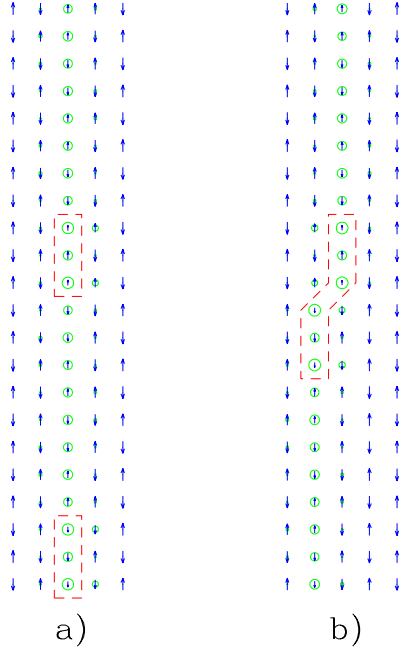}}
\end{center}
\caption{
Addition of a single hole to a half filled stripe. Hartree-Fock spin and
charge textures calculated for a $5 \times 22$ 
lattice with periodic boundary conditions and 12 holes
($U/t = 8$). a) Two kink solution. b) Step solution.}
\label{fig:solitons}
\end{figure}

Upon doping, we expect that each electron
leads to two charge 1/2 solitons\cite{G93}. The Hartree-Fock method is well
suited to visualize charge fractionalization of this type\cite{HKSS88},
as shown in Fig.[\ref{fig:solitons} a)]. The fact that the stripe
is embedded in a two dimensional background allows for an
alternative solution, the step\cite{BSZ00}, also shown in
Fig.[\ref{fig:solitons} b)]. This solution has a slightly lower
energy. Thus, we expect that deviations from half filling
in the stripes will lead to the formation of steps. 
We can calculate the quantum dynamics of the steps along the
direction parallel to the stripe. Within the CI method, a step
like the one shown in Fig..[\ref{fig:solitons} b)] gives rise to
a band of delocalized states of width $0.59 t \, ( U/t = 8)$
and $0.37 t \, ( U/t = 12)$. This bandwidth measures directly
the delocalization of the step along the stripe due to quantum
fluctuations. The 
energy gain due to this hybridization, $0.29 t \, ( U/t = 8 )$ and
$0.18 t \, ( U/t = 12$),   further stabilizes
the step.  
The formation of steps is hindered when the stripes are sufficiently close.
Experiments suggest that the closest distance between stripes is four
lattice units, which is realized at 1/8 doping\cite{Zetal99}.
Near this filling, holes in the stripes will give rise to charge
fractionalization and Luttinger liquid behavior\cite{CG98,CCR99}.
   
\begin{figure}
\mbox{\epsfxsize 8cm \epsfbox{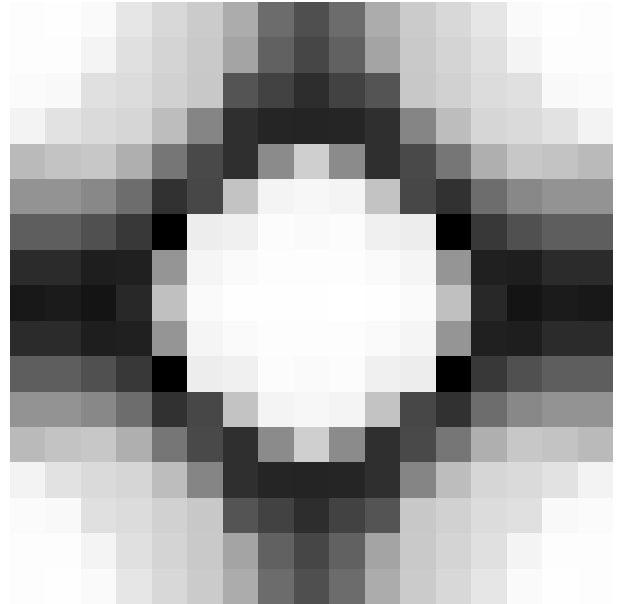}}
\caption{Values of $n_{\bf \vec{k}}$ in a $16 \times 16$ cluster with
periodic boundary conditions.}
\label{fig:nk}
\end{figure}

We have also computed the value of $n_{\bf \vec{k}}$, from the solutions
described above.
They are shown in fig.[\ref{fig:nk}], calculated in a $16 \times 16$
cluster, and obtained by integrating in energies
over the top half of the valence band and the stripe states
induced within the gap. 
Note that the charge and spin distributions are homogeneous
in real space, as the CI method restores full translational symmetry.
The strong anisotropy, in ${\bf \vec{k}}$ space, of $n_{\bf \vec{k}}$
reflects the existence of the stripe (see also\cite{IM99,Zetal00}), 
and is consistent with
the experimental results reported in\cite{Zetal99}.
The usefulness of the method used here for the study of
half filled stripes suggests that a detailed description of the spin waves in the
antiferromagnetic background is not essential\cite{SDW}. On the other hand, 
the scheme used here shows that there is a delicate balance between spin
polarons\cite{VL91,LG98} and stripes, 
which seems also to be found in Montecarlo
calculations of the t-J model\cite{WS98,HM99,CS99,WS00,Metal00}. 
A detailed comparison of
the  relative stability of these phases cannot be done within the present
method, and seems also difficult to achieve by other
techniques. It would be interesting to analyze further the experimental
implications of this near degeneracy of qualitatively different solutions.
Finally, the present method suggests that inhomogeneous spin and charge
textures, like stripes or spin polarons, can be viewed as microsocopic
manifestations of phase separation, as the natural  homogeneous 
solution which can be obtained in mean field has negative compressibility\cite{Getal00}.

In conclusion, we show that the  half filled stripes observed in experiments
in cuprates can be derived using the same methods which lead to the prediction
of the existence of (filled) stripes 
in the Hubbard model\cite{ZG89,PR89,Sc90,VL91}.
The method allows us to
describe the quantum transverse fluctuations of the stripes, which are
significant (comparable to $t$) for segments of length
of 4-6 lattice units. 
Finally, we present the distribution
of $n_{\bf \vec{k}}$, with good agreement with experimental results.

The financial support of the CICyT (Spain), through
grants no. PB96-0875, PB96-0085
and 1FD97-1358 and CAM (Madrid), through grant
no. 07N/0045/98 is gratefully acknowledged.

\end{document}